\documentclass[english]{article}
\usepackage[T1]{fontenc}
\usepackage[latin9]{inputenc}
\usepackage{amsmath}
\usepackage{amssymb}
\usepackage{graphicx}

\makeatletter

\usepackage{graphics}

\newcommand{\mathsym}[1]{{}}

\makeatother

\usepackage{babel}
\begin{document}

\title{Dirac Field in FRW Spacetime: Current and Energy Momentum}

\author{P.R. Dhungel and U. Khanal \\
 Central Department of Physics, Tribhuvan University, Kirtipur, \textbf{Nepal}}

\date{\today}
\maketitle
\begin{abstract}
The behaviour of the Dirac field in FRW space-time is investigated.
The relevant equations are solved to determine the particle and energy
distribution. The angular and radial parts are solved in terms of
Jacobi polynomials. The time dependence of the massive field is solved
in terms of known function only for the radiation filled flat space.
WKB method is used for approximate solution in general FRW space.
Of the two independent solutions, one is found to decay in time as
the Universe expands, while the other solution grows. This could be
the source of the local particle current. The behaviour of the particle
number and energy density are also investigated. It is found that
the particles arrange themselves in a number and density distribution
pattern that produces a constant Newtonian potential as required for
the flat rotation curves of galaxies. Further, density contrast is
found to grow with the expansion.
\end{abstract}
PACS: 03.65.Pm, 04.20.Cv, 04.98.80.Jk

Keywords: Dirac equation, NP formalism, FRW space-time, Spinors.

\section{Introduction}

Since all matter is ultimately fermionic, behaviour of Dirac field
in expanding Universe must reveal hitherto unknown features of the
Universe. The properties of matter fields in general, and the massive
Dirac fields in particular, must have profound consequences on the
structure formation proecess. 

The spinor formalism developed by B. v. d. Waerden\cite{Waerden}
in 1928 and later elaborated by O. Laporte, et al comprises all representations
of the Lorentz group, even those not contained in ordinary tensor
analysis. However, it was not widely used at that time. Later, in
1936, the spinor calculus was used by Dirac\cite{Dirac-1936} to derive
wave equations for particles with spin greater than one half, that
means wave equations for (elementary) particles that had not been
discovered up to then. This made the spinor calculus known to a wider
audience and Brill and Wheeler's \cite{Brill-Wheeler-1957}analysis
and application to the interaction of neutrinos with gravitational
fields further added weights to this. When Newman and Penrose developed
a special formalism\cite{Newman-Penrose-1962} of projecting vectors,
tensors and spinors onto a set of null tetrad bases (explained in
detail in Ref. \cite{Chandra}), it became a very useful tool to study
the behaviour of quantum-fields of all spins in general relativistic
back-ground gravity. This method has been used successfully in various
black hole geometries \cite{Chandra,Lohia and Panchapakesan,UK and Panchapakesan,uk3}
and more recently to further study the behaviour of Dirac particles
in different geometries \cite{Haghighipour-2005,Sharif-2002,Bicak,Badawi and Sakalli,Zecca-2009}. 

In continuance of previous work \cite{uk1,uk2} on applying this method
to investigate the behaviour of the fields of electrodynamics in Friedmann-Robertson-Walker(FRW)
space-time, we take up the Dirac field to get some insight on the
process of structure formation. For our purpose, we first write the
metric as \cite{kolb,wein}

\begin{equation}
ds^{2}=a^{2}\left[d\eta^{2}-dr^{2}-S^{2}\left(d\theta^{2}+\sin^{2}\theta d\phi^{2}\right)\right]
\end{equation}

where $a$ is the scale factor that depends on the conformal time
$\eta$ which is related to the cosmic co-moving time $t$ by $dt=ad\eta$,
and

\begin{equation}
\begin{array}{cc}
S=\frac{\sin\left(\sqrt{K}r\right)}{\sqrt{K}}= & \begin{cases}
\sin r, & K=1\\
r, & K=0\\
\sinh r, & K=-1
\end{cases}\end{array}.\label{eq:S in FRW metric}
\end{equation}

The flat case can be considered to be the limit of either the closed
or open cases for $K\rightarrow0$. Thus, in the complexified $r\mbox{-plane}$,
the closed Universe lies along the real axis, open along the imaginary
and flat near the origin. So, we may consider the proposition that,
as an overall flat space expands, it fragments into under-density
voids that develop locally as an open FRW space along the imaginary
axis, and the overdensity regions evolve locally into the large scale
structures as the closed counterpart. We choose the null tetrad as
$l_{\mu}=[1,-1,\,0,\,0]$, $n_{\mu}=\frac{a^{2}}{2}[1,\,1,\,0,\,0]$,
$m_{\mu}=-\frac{aS}{\sqrt{2}}[0,\,0,\,1,\, i\sin\theta]$ and the
complex conjugate $\bar{m}_{\mu}$so that the directional derivatives
can be expressed as 

\begin{eqnarray}
D=l^{\mu}\partial_{\mu} & =\frac{1}{a^{2}}\mathcal{D}_{-}, & \Delta=n^{\mu}\partial_{\mu}-\frac{1}{2}\mathcal{D}_{+}\nonumber \\
\delta=m^{\mu}\partial_{\mu} & =\frac{1}{\sqrt{2}aS}\mathcal{L}_{-}, & \delta^{*}=\bar{m}^{\mu}\partial_{\mu}=\frac{1}{\sqrt{2}aS}\mathcal{L}_{+}
\end{eqnarray}

where,

\begin{eqnarray}
\mathcal{D}_{\pm} & = & \frac{\partial}{\partial r}\mp\frac{\partial}{\partial\eta}
\end{eqnarray}
and 
\begin{equation}
\mathcal{L}_{\pm}=\frac{\partial}{\partial\theta}\mp\frac{i}{\text{sin\ensuremath{\theta}}}\frac{\partial}{\partial\phi}
\end{equation}

The spinor equivalent of these directional derivatives are 

\begin{equation}
\partial_{00'}=D,\,\,\,\,\,\partial_{11'}=\Delta,\,\,\,\,\,\partial_{01'}=\delta,\,\,\,\,\, and\,\,\,\,\,\partial_{10'}=\delta^{*}\label{eq:Spinor-dir-derivatives}
\end{equation}

Also, the non-vanishing spin coefficients \cite{Chandra} are given
by $\beta=-\alpha=\frac{\text{cot\ensuremath{\theta}}}{2\sqrt{2}S},\,\gamma=-\frac{1}{2a}\frac{\partial a}{\partial\eta},\,2\mu=\frac{1}{a}\frac{\partial a}{\partial\eta}-\frac{1}{S}\frac{\partial S}{\partial r},\,-a^{2}\rho=\frac{1}{a}\frac{\partial a}{\partial\eta}+\frac{1}{S}\frac{\partial S}{\partial r}$.

Dirac equations for spin-1/2 particles represented by a pair of spinors,
$P^{A}$~~and~ $\bar{Q}^{A'}$, are 

\begin{equation}
\sigma_{\,\,\, AB'}^{i}\partial_{i}P^{A}+i\frac{M}{\sqrt{2}}\bar{Q}_{B'}=0
\end{equation}

and

\begin{equation}
\sigma_{\,\,\, AB'}^{i}\partial_{i}Q^{A}+i\frac{M}{\sqrt{2}}\bar{P}_{B'}=0,
\end{equation}

where $\sigma_{\,\,\, AB'}^{i}$~~are Pauli-martices and M is the
mass of the particle (expressed as the inverse of its Compton wave-length).
To write these equations in the Newman-Penrose formalism in a curved
space-time, we replace the ordindary derivatives with covariant derivatives
and Pauli-matrices by $\sigma$ generalized to curved spacetime as

\begin{equation}
\sigma_{\,\,\, AB'}^{i}=\frac{1}{\sqrt{2}}\left|\begin{array}{cc}
l^{i} & m^{i}\\
\bar{m}^{i} & n^{i}
\end{array}\right|
\end{equation}

Thus the Dirac equations are

\begin{equation}
\sigma_{\,\,\, AB'}^{i}P_{\,\,\,;\, i}^{A}+i\frac{M}{\sqrt{2}}\bar{Q}^{C'}\varepsilon_{C'B'}=0\label{eq:DiracEq-1}
\end{equation}

and

\begin{equation}
\sigma_{\,\,\, AB'}^{i}Q_{\,\,\,;\, i}^{A}+i\frac{M}{\sqrt{2}}\bar{P}^{C'}\varepsilon_{C'B'}=0,\label{eq:DiracEq-2}
\end{equation}

where

\begin{equation}
\varepsilon_{C'B'}=\left|\begin{array}{cc}
\,\,0 & 1\\
-1 & 0
\end{array}\right|
\end{equation}

In order to write explicit forms of these equations in terms of spin
coefficients, consider the case $B'=0$ in Eq. (\ref{eq:DiracEq-1}):

\begin{equation}
\sigma_{\,\,\,00'}^{i}P_{\,\,\,;\, i}^{0}+\sigma_{\,\,\,10'}^{i}P_{\,\,\,;\, i}^{1}-i\frac{M}{\sqrt{2}}\bar{Q}^{1'}=0,
\end{equation}

Explicitly, in NP formalism this becomes

\begin{equation}
(\partial_{00'}P^{0}+\Gamma_{\,\,\, d00'}^{0}P^{d})+(\partial_{10'}P^{1}+\Gamma_{\,\,\, d10'}^{1}P^{d})=i\frac{M}{\sqrt{2}}\bar{Q}^{1'}
\end{equation}

Noting that $\Gamma^{1}=\varepsilon^{10}\Gamma_{0}=-\Gamma_{0}$ and
$\Gamma^{0}=\varepsilon^{01}\Gamma_{1}=\Gamma_{1}$and using (\ref{eq:Spinor-dir-derivatives}),
this equation can be written as

\begin{equation}
DP^{0}+(\Gamma_{1000'}P^{0}+\Gamma_{1100'}P^{1})+\delta^{*}P^{1}-(\Gamma_{0010'}P^{0}+\Gamma_{0110'}P^{1})=i\frac{M}{\sqrt{2}}\bar{Q}^{1'}
\end{equation}

Replacing the various spin coefficient by their special symbols \cite{Chandra},
we obtain

\begin{equation}
(D+\varepsilon-\rho)P^{0}+(\delta^{*}+\pi-\alpha)P^{1}=i\frac{M}{\sqrt{2}}\bar{Q}^{1'}
\end{equation}

Similarly, for $B'=1$ in Eq. (\ref{eq:DiracEq-1}) gives another
such equation and following the similar steps with the complex conjugate
of Eq. (\ref{eq:DiracEq-2}) gives the other two equations. Hence
the Dirac equations for the components of the Dirac spinors in NP
formalism are given by

\begin{align}
(D+\varepsilon-\rho)P^{0}+(\delta^{*}+\pi-\alpha)P^{1} & =i\frac{M}{\sqrt{2}}\bar{Q}^{1'},\nonumber \\
(\Delta+\mu-\gamma)P^{1}+(\delta+\beta-\tau)P^{0} & =-i\frac{M}{\sqrt{2}}\bar{Q}^{0'},\nonumber \\
(D+\varepsilon^{*}-\rho^{*})\bar{Q}^{0'}+(\delta+\pi^{*}-\alpha^{*})\bar{Q}^{1'} & =-i\frac{M}{\sqrt{2}}P^{1},\,\,\, and\nonumber \\
(\Delta+\mu^{*}-\gamma^{*})\bar{Q}^{1'}+(\delta^{*}+\beta^{*}-\tau^{*})\bar{Q}^{0'} & =i\frac{M}{\sqrt{2}}P^{0}.\label{eq:4DiracEqs}
\end{align}

In the next section, the solution of these equations including angular,
radial and time parts are presented. Section 3 deals with the particle
current and the energy-momentum of the field. Concluding remarks are
made in the last section. Although the solutions of Dirac equations
have been obtained earlier \cite{Barut-Duru-1987,Zecca-1998,Zecca-2006,Huang-2005},
they are either restricted to some special cases or the expressions
are comparatively complex ones. With our choice of null tetrad bases,
the separated equations are more tractable and the solutions can be
written in terms of known mathematical functions.

\section{Solutions}

Substituting ($\sqrt{2}SaP^{0},\, Sa^{2}P^{1},\,-\sqrt{2}Sa\bar{Q}^{0'},\, Sa^{2}\bar{Q}^{1'}$)
with

($\phi_{-1/2}Y_{-1/2},\,\phi_{1/2}Y_{1/2},\,\phi_{-1/2}Y_{1/2},\,\phi_{1/2}Y_{-1/2}$)
in the Eqs. (\ref{eq:4DiracEqs}), using the directional derivatives
and spin-coefficients given in Sec. 1, the angular parts of Eqs. (\ref{eq:4DiracEqs})
are found to separate into 

\begin{equation}
\left(\mathcal{L}_{\pm}+\frac{1}{2}\cot\theta\right)Y_{\pm1/2}=\mp\lambda Y_{\mp1/2},\label{eq:Dirac-angular}
\end{equation}

while the radial and time parts satisfy

\begin{equation}
r\mathcal{D}_{\pm}\phi_{\pm1/2}=\left(f\lambda\mp iMra\right)\phi_{\mp1/2}\label{eq:Dirac-RadialTime}
\end{equation}

Substituting for $Y_{+1/2}=\left(\mathcal{L}_{-}+\frac{1}{2}\cot\theta\right)Y_{-1/2}$from
Eq. (\ref{eq:Dirac-angular}) into the one for $Y_{-1/2}$, and a
similar process with $Y_{-1/2}$, gives us the two angular equations

\begin{equation}
\left(\frac{1}{\sin\theta}\frac{\partial}{\partial\theta}\sin\theta\frac{\partial}{\partial\theta}\pm\frac{i\cos\theta}{\sin^{2}\theta}\frac{\partial^{2}}{\partial\varphi^{2}}-\frac{1}{4}\cot^{2}\theta-\frac{1}{2}\lambda^{2}\right)Y_{\pm1/2}=0
\end{equation}

Explicitly, the angular parts representing the two helicities are
solved in terms of the Jacobi polynomials $P_{n}^{(\alpha,\beta)}$
given by the spin weighted spherical harmonics

\begin{equation}
_{\pm s}Y_{l}^{m}(\theta,\phi)=Ne^{im\phi}(1-\cos\theta)^{\frac{m}{2}\pm\frac{s}{2}}(1+\cos\theta)^{\frac{m}{2}\mp\frac{s}{2}}P_{l-m}^{(m\pm s,m\mp s)}(\cos\theta)
\end{equation}
 of spin weight $s=\frac{1}{2}$; here, the total angular momentum
$l=s+1,\, s+2,\,...$, are half-integers as they already include the
spin and the separation constant $\lambda=(l+1/2)$. $_{\pm s}Y_{l}^{m}$
are appropriately normalized and weighted to give $\int_{\theta=0}^{\pi}\int_{\phi=0}^{2\pi}\,_{s}Y_{l^{\prime}}^{m^{\prime}*}{}_{s}Y_{m}^{l}\sin\theta d\theta d\phi$$=\delta_{ll^{\prime}}\delta_{mm^{\prime}}$.
The radial and temporal parts can also be separated by writing $R_{\pm}T_{\pm}=\phi_{1/2}\pm\phi_{-1/2}$
in Eq. (\ref{eq:Dirac-RadialTime}) to give the equations

\begin{equation}
\frac{1}{R_{\mp}}\left(\frac{\partial}{\partial r}\mp\frac{l+1/2}{S(r)}\right)R_{\pm}=\frac{1}{T_{\pm}}\left(\frac{\partial}{\partial\eta}\pm iMa\right)T_{\mp}=ik.\label{1.1}
\end{equation}
The separation constant $k=pa$ can be identified with the comoving
momentum. The solutions labeled $\pm$ can be decoupled easily\cite{uk1},
and we believe that they represent the negative and positive energy
solutions respectively. The decoupled equatins are of second order
and are given by 

\begin{equation}
\left(\frac{\partial}{\partial r}\pm\frac{l+1/2}{S(r)}\right)\left(\frac{\partial}{\partial r}\mp\frac{l+1/2}{S(r)}\right)R_{\pm}=-k^{2}R_{\pm}
\end{equation}

and

\begin{equation}
\left(\frac{\partial}{\partial\eta}\pm iMa\right)\left(\frac{\partial}{\partial\eta}\mp iMa\right)T_{\pm}=-k^{2}T_{\pm}.
\end{equation}

\subsection{Radial Solutions}

\begin{figure}
\includegraphics[scale=1.5]{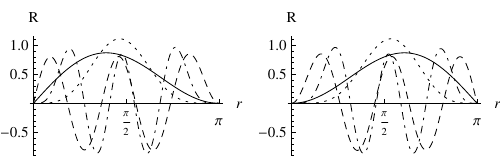} \\
\caption{The radial dependence, $R{}_{+}$ on the right and $R_{-}/i$ on the
left. The $l=1/2,$ are filled for $k=3/2\mbox{ and dotted for }5/2$
respectively; the $l=7/2$ are dashed for $k=11/2\mbox{ and dot-dashed for}k=17/2$. }

\label{radial} 
\end{figure}

The resulting regular radial solutions 
\begin{eqnarray}
R_{\pm} & = & A_{\pm}\left(1-\cos\sqrt{K}r\right)^{\frac{l+1}{2}\mp\frac{1}{4}}\left(1+\cos\sqrt{K}r\right)^{\frac{l+1}{2}\pm\frac{1}{4}}\label{1.2}\\
 &  & P_{\frac{k}{\sqrt{K}}-l-1}^{\left(l+\frac{1}{2}\mp\frac{1}{2},\frac{l}{2}+\frac{1}{2}\pm\frac{1}{2}\right)}\left(\cos\sqrt{K}r\right),\nonumber 
\end{eqnarray}
 are the appropriately weighted Jacobi polynomials that can be normalized
to give $\int_{0}^{\pi}\left|R_{\pm}\right|^{2}dr=1$. The behaviour
of $R_{\pm}$ is displayed in Fig. (\ref{radial}). It suffices to
write the radial solutions for all $K^{\prime}s$ in the form of Eq.(\ref{1.2}),
as the limit to flat space is achieved with $K\rightarrow0$, in which
case the asymptotic behaviour of the Jacobi polynomial of large order
leads to \cite{abram} 

\begin{equation}
R_{\pm}=A_{\pm}rj_{l\mp\frac{1}{2}}(kr)\label{eq:R for flat}
\end{equation}

where $j_{n}$ are the spherical Bessel functions. It was shown in
Ref. \cite{uk1} that $R_{+}^{\ast}R_{-}=-R_{+}R_{-}^{\ast}$, so
that $A_{-}=iA_{+}$. As the mass does not appear in the radial solution,
the spatial solution of the massive and massless field equations are
the same. The mass couples with the scale factor and appears only
in the temporal part. One important aspect of the solution Eq. (\ref{1.2})
is that $k-l-1=n$ has to be an integer for $R_{\pm}$ to be regular
at $r=0$ and $\pi$. As $l\geq\frac{1}{2}$ is a half-integer increasing
in steps of unity from its minimum value, $k\geq\frac{3}{2}$ is also
half-integer. Hence, the momentum of the free Dirac field in closed
FRW space-time is quantized in steps of unity from $\frac{3}{2}$
upwards. This minimum value of $\frac{3}{2}$ can be interpreted as
half unit of spin and one unit representing the mass in the lowest
momentum state for the massive or one unit of lightlike momentum for
the massless fields. Eq (\ref{1.1}) leads to a number of Wronskian-like
conditions, the most important of which that we will require later
are~ 
\begin{align}
\left|T_{+}\right|^{2}+\left|T_{-}\right|^{2} & =1\nonumber \\
ik\left[T_{+}^{\ast}T_{-}-T_{+}T_{-}^{\ast}\right] & =\frac{d}{d\eta}\left|T_{+}\right|^{2}\label{1.3}\\
ik\left[\left|R_{+}\right|^{2}-\left|R_{-}\right|^{2}\right] & =\frac{d}{dr}R_{+}^{\ast}R_{-}=-\frac{d}{dr}R_{+}R_{-}^{\ast}.\nonumber 
\end{align}

\subsection{Time Evolution}

The decoupled temporal equation is 
\begin{equation}
\left(\frac{d^{2}}{d\eta^{2}}+k^{2}+M^{2}a^{2}\mp iMa^{\prime}\right)T_{\pm}=0,\label{2.1}
\end{equation}
 and $a^{\prime2}=\left(\frac{da}{d\eta}\right)^{2}=H_{0}^{2}\left[\Omega_{\Lambda}a^{4}+\left(1-\Omega_{\Lambda}-\Omega_{M}-\Omega_{R}\right)a^{2}+\Omega_{M}a+\Omega_{R}\right]$
is given by the Friedmann-Le Maitre equation with $\Omega_{\Lambda},\,\Omega_{M}\,$and
$\Omega_{R}$ being the densities contributed by the cosmological
constant, matter and radiation respectively, in units of the critical
density at the time $\eta_{0}$ when the scale factor is normalized
to $a_{0}=1$. Given a functional form for $a\left(\eta\right)$,
it is always possible to write a series solution\cite{uk2} of Eq.(\ref{2.1}).
It is instructive to start with solutions for the massless field.
The two solutions are just 
\begin{equation}
T_{\pm}=\pm\frac{1}{\sqrt{2}}e^{-ik\eta},\label{2.2a}
\end{equation}
 where the multiplicative constants and their signs have been fixed
by the conditions imposed by Eqs. (\ref{1.1}) and (\ref{1.3}).

\begin{figure}
\includegraphics{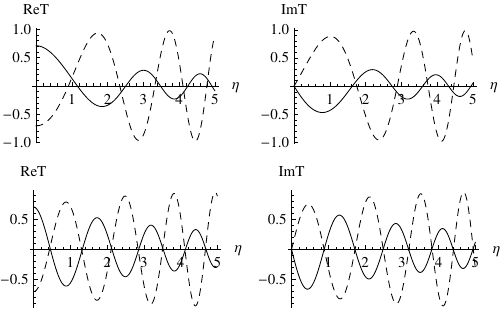} \\
\caption{The temporal dependence of the Dirac field in a radiation filled flat
FRW space-time. $T_{+}$ are shown by solid and $T_{-}$ by dashed
lines respectively. The upper two are real and imaginary parts for
$\frac{k}{\sqrt{H_{0}M}}=\frac{3}{2}$, and the lower two for $\frac{k}{\sqrt{H_{0}M}}=\frac{7}{2}$. }

\label{tplmin} 
\end{figure}

\begin{figure}
\includegraphics{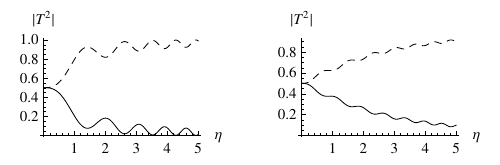} \\
\caption{Solid curves are$\left|T_{+}\right|^{2}$, and dashed are$\left|T_{-}\right|^{2}$;
$\frac{k}{\sqrt{H_{0}M}}=\frac{3}{2}$ are at the left and $\frac{k}{\sqrt{H_{0}M}}=\frac{7}{2}$
at the right. $\left|T_{+}\right|^{2}$ are seen to decay to zero
while$\left|T_{-}\right|^{2}$ grow to unity. }

\label{t2} 
\end{figure}

As shown in Ref. \cite{uk1}, the time dependence can be solved in
terms of a known special function (Whitaker) only in the case of the
radiation filled flat Universe. This case is very important for the
reason that it represents the nature of the early radiation dominated
Universe from the very big-bang during which time, the density parameter
was extremely fine-tuned to almost unity. So before trying to approximate
the general solution, let us consider this in some further details.
The solutions are better represented in terms of the parabolic cylinder
functions\cite{grads} as 
\begin{eqnarray}
T_{\pm}\left(\eta\right) & = & B_{\pm}D_{\pm\frac{ik^{2}}{2MH_{0}}}\left[\left(1\mp i\right)\sqrt{MH_{0}}\eta\right]\label{2.2}\\
 &  & \mp\frac{k}{2\sqrt{MH_{0}}}\left(1\pm i\right)B_{\mp}D_{\mp\frac{ik^{2}}{2MH_{0}}-1}\left[\left(1\pm i\right)\sqrt{MH_{0}}\eta\right]\nonumber 
\end{eqnarray}
 where Eq.(\ref{1.1}) has been used to relate the constants of the
two independent solutions. $\sqrt{MH_{0}}$ is the reciprocal of the
geometric mean of the Hubble time of the Universe and the Compton
time of the Dirac particle. From the discussions of the previous paragraph,
we can fix the constants to be 
\[
B_{\pm}=\pm\frac{1}{\sqrt{2}}e^{-\frac{\pi k^{2}}{4H_{0}M}}\left[D_{\mp\frac{ik^{2}}{2MH_{0}}}\left(0\right)\mp\frac{k\left(1\pm i\right)}{2\sqrt{H_{0}M}}D_{\mp\frac{ik^{2}}{2MH_{0}}-1}\left(0\right)\right]
\]
 by requiring $T_{\pm}\left(0\right)=\pm\frac{1}{\sqrt{2}}$. The
behaviour of the solutions Eq.(\ref{2.2})\ for some values of $k$
are shown in Fig. (\ref{tplmin} and \ref{t2}).

Next, we try to approximate the solutions of Eq. (\ref{2.2}) by WKB
method, assuming that $a$ changes very slowly with $\eta$ so that
$a^{\prime}\left(\eta\right)$ is small, i. e. the expansion rate
of the Universe is small compared to the size of the Universe. Multiplying
the derivative in Eq.(\ref{1.1}) with $\delta$ to incorporate the
order of the approximation and substituting $T_{\pm}=\exp\left(-\frac{i}{\delta}\sum\limits _{n=0}^{\infty}\delta^{n}\int f_{n}d\eta\right)$
in Eq. (\ref{2.1}) we get the equation for $f$: 
\begin{equation}
\sum\limits _{n=0}^{\infty}\delta^{n}\left\{ \sum\limits _{m=0}^{n}\left(f_{m}f_{n-m}+if_{n-1}^{\prime}\right)\right\} =k^{2}+M^{2}a^{2}\mp i\delta Ma^{\prime}.\label{2.3}
\end{equation}
 Equating equal powers of $\delta$ we get the equations for the respective
orders of the approximations: 
\begin{eqnarray}
f_{0} & = & \sqrt{k^{2}+M^{2}a^{2}}=a\varepsilon_{0},\nonumber \\
2f_{0}f_{1} & = & -i\left(f_{0}^{\prime}\pm Ma^{\prime}\right)\text{ and}\label{2.4}\\
-if_{n-1}^{\prime} & = & \sum\limits _{m=0}^{n}f_{m}f_{n-m}.\nonumber 
\end{eqnarray}
 Solving to the first order, we find $\int f_{1}d\eta=i\ln\sqrt{1\mp\frac{M}{\varepsilon_{0}}}$,
so we can write the solution to first order as 
\begin{equation}
T_{\pm}=\pm\sqrt{\frac{\varepsilon_{0}\mp M}{2\varepsilon_{0}}}e^{-i\int a\varepsilon_{0}d\eta},\label{2.5}
\end{equation}
 which resembles the solution in Minkowskian space-time. The solution
also gives us another important information, viz., at late times (in
the evolution of the Universe) when $a\rightarrow\infty,\,\varepsilon_{0}\rightarrow M$,
so $T_{+}\rightarrow0$ and $T_{-}\rightarrow-1$. This shows that
one of the solutions decays in time as the Universe expands, while
the other solution grows. So, we suspect that the former one should
be negative energy solution and the latter the positive energy solution
because it is in consistent with the behaviors of the Dirac field
in the Minkowski space where it is found that the negative energy
solutions exist only in the relativistic regime while in the non-relativistic
limit it is the positive energy solution that dominates \cite{Roman-1965}.
However, it has to be confirmed with further investigations.

\section{Particle Current and Energy-Momentum}

\begin{figure}
\includegraphics[scale=1.5]{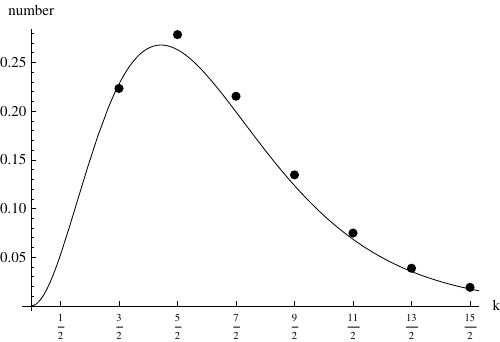} \\
\caption{The relative particle number distribution of extremely relativistic
fermions in closed FRW space. The momentum $k$ is quantized upwards
in steps of unity from the minimum value of $3/2$, and the filled
dots represent the relative number of particles with that momentum.
The filled curve is the Fermi-Dirac distribution. We see that at low
momenta, the two distributions are quite different, but for high $k$ the
two start to coincide. }

\label{fddist} 
\end{figure}

The $\eta$ and $r$ components of the four-current integrated over
the angular co-ordinates are found to be (with S defined in Eq. (\ref{eq:S in FRW metric}))
\begin{eqnarray}
4\pi a^{4}J^{\eta}S^{2} & = & 4\pi na^{3}S^{2}=\left[\left|R_{+}\right|^{2}\left|T_{+}\right|^{2}+\left|R_{-}\right|^{2}\left|T_{-}\right|^{2}\right]=\text{ 4\ensuremath{\pi S^{3}a^{3}\left(n_{-}+n_{+}\right)}}\nonumber \\
\text{ and } &  & 4\pi a^{4}J^{r}S^{2}=-R_{+}^{\ast}R_{-}\left(T_{+}^{\ast}T_{-}-T_{+}T_{-}^{\ast}\right)=-\frac{R_{+}^{\ast}R_{-}}{ik}\frac{d}{d\eta}\left|T_{+}\right|^{2}.\label{3.1a}
\end{eqnarray}
 At this point, we are not much concerned with the other components,
viz. $J^{\theta}\propto Y_{+}^{\ast}Y_{-}-Y_{+}Y_{-}^{\ast}=0$ and
$a^{4}J^{\phi}S^{2}=-\frac{iR_{+}^{\ast}R_{-}}{2\sin\theta}\left(T_{+}^{\ast}T_{-}+T_{+}T_{-}^{\ast}\right)\left(Y_{+}^{\ast}Y_{-}+Y_{+}Y_{-}^{\ast}\right)$
which is independent of $\phi.$ The equation of continuity becomes
$\frac{\partial}{\partial\eta}na^{3}S^{2}=-\frac{\partial}{\partial r}a^{4}S^{2}J^{r}=$
$\frac{1}{ik}\frac{dR_{+}^{\ast}R_{-}}{dr}\frac{d\left|T_{+}\right|^{2}}{d\eta}$
on using the relations Eq. (\ref{1.3}). Integrating the comoving
number density over the whole volume, the total number contained within
the whole of the closed Universe gives us the normalization condition
$N_{T}a^{3}=\int\nolimits _{0}^{\pi}dr\int na^{3}\sin^{2}rd\Omega=\left|T_{+}\right|^{2}+\left|T_{-}\right|^{2}=1$
so that the total comoving number is conserved. In the extreme relativistic
regime, i.e., in very early epoch of the Universe, when its size was
small ($a\rightarrow\mbox{small}$), assuming a Fermi-Dirac(FD) type
of distribution, and taking the volume of the closed space to be $V=\intop_{0}^{\pi}S^{2}dr\int d\Omega=2\pi^{2}$,
we may write the average number density of the particles as
\begin{equation}
na^{3}=\frac{1}{2\pi^{2}}\sum_{k,l,m}\frac{1}{e^{k+l+1}+1}=\frac{1}{2\pi^{2}}\sum_{k=0}^{\infty}\frac{\left(k+1\right)\left(k+2\right)}{e^{k+3/2}+1}=0.0827975.\label{3.1b}
\end{equation}
The FD value is $na^{3}=0.0913454$, and this distribution deviates
significantly from Fermi-Dirac at low $n$. The number distribution,
corresponding to the massless case, is shown and compared with the
FD distribution in Fig. (\ref{fddist}).

On the other hand, integrating over the angular co-ordinates and only
out to a radius $r$, we have 
\begin{eqnarray}
Na^{3} & = & \int_{0}^{r}na^{3}S^{2}\, dr\, d\Omega=\int_{0}^{r}\left|R_{-}\right|^{2}dr+\frac{R_{+}^{\ast}R_{-}}{ik}\left|T_{+}\right|^{2}\nonumber \\
 & \rightarrow & \int_{0}^{r}\left|R_{-}\right|^{2}dr+\frac{R_{+}^{\ast}R_{-}}{2ik}\left(1-\frac{Ma}{\sqrt{k^{2}+M^{2}a^{2}}}\right),\label{3.2}\\
 &  & \text{1}^{\text{st}}\,\text{order}\,\text{WKB}\nonumber 
\end{eqnarray}
 for the relative number of particles within a finite volume. This
WKB result are plotted for representative cases in Fig. (\ref{numbercl}).

\begin{figure}
\includegraphics{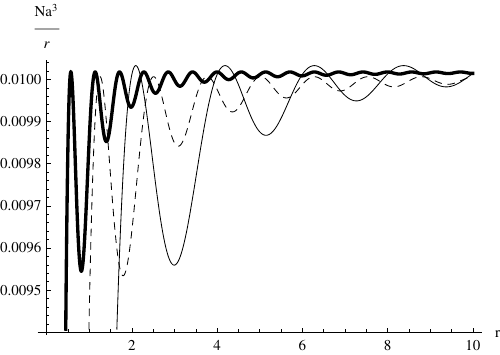}

\caption{\label{fig:Na3-by-r-t0}The comoving particle number ($Na^{3}/r)$
for flat Universe at $\eta\rightarrow0$. The normal solid line, dashed
line and thick solid line are for k = 3/2, 5/2 and 11/2 respectively,
all of which are for $l$ = 1/2. They are appropriately scaled for
comparision purpose. The quantities are found to saturate to a constant
value resembling to the flattening of the rotation curves of galaxies.}
\end{figure}

\begin{figure}
\includegraphics{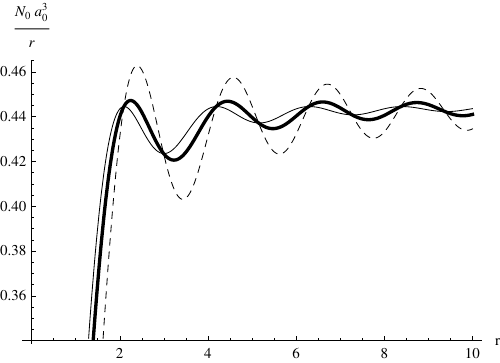}

\caption{\label{fig:Na3-by-r-M varied}The comoving particle number ($Na^{3}/r$)
for massive flat FRW spacetime with $l$ = 1/2 and k = 3/2. The normal
solid, dashed and thick solid curves are for M =0.01, 1 and 100 respectively.
The behaviour is same as in the case of massless field except that
amplitude of fluctuations are larger, and increase for increasing
mass. The graphs for the three cases have been appropriately scaled
to make of almost equal heights for comparision.}
\end{figure}

\begin{figure}
\includegraphics{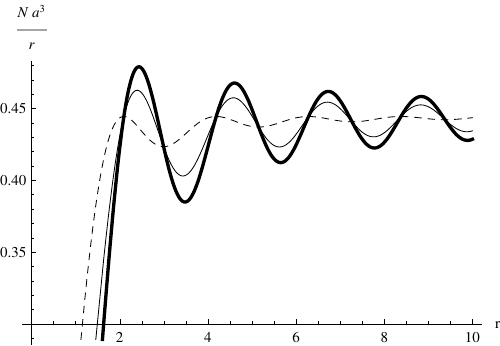}

\caption{\label{fig:Na3-by-r-t varied}Evolution of $Na^{3}/r$ at different
conformal times $\eta$ for M = 1. The normal solid curve is for $\sqrt{MH_{0}}\eta=1$,
the dashed curve is for $\sqrt{MH_{0}}\eta=.01$ and the thick solid
curve is for $\sqrt{MH_{0}}\eta=10,000$. All the graphs are for \textit{l}
= 1/2 and k = 3/2. It is seen that the density contrast increases
with the elapse of time.}
\end{figure}

Particularly, for flat Universe, the comoving particle number and
the particle current can be obtained using the expression for $R_{\pm}$~~given
by Eq. (\ref{eq:R for flat}) and $T_{\pm}$~~given by Eq. (\ref{2.2}).
Their sample plots have been shown in Figures (\ref{fig:Na3-by-r-t0}),
(\ref{fig:Na3-by-r-M varied}) (with different values of M) and (\ref{fig:Na3-by-r-t varied})
(with different values of conformal time~ $\eta$).

Next, we consider the energy-momentum tensor. At present we are concerned
only with the $\eta\eta$ and $rr$ components giving the density
$\rho$ and the pressure $P_{r}$ respectively. After integration
over the angular co-ordinates, we end up with 
\begin{eqnarray}
4\pi S^{2}\rho a^{4} & = & -\frac{k}{2}\left(\left|R_{+}\right|^{2}+\left|R_{-}\right|^{2}\right)\left(T_{+}^{\ast}T_{-}+T_{+}T_{-}^{\ast}\right)\nonumber \\
 &  & -\left[Ma\left(\left|Z_{+}\right|^{2}-\left|Z_{-}\right|^{2}\right)\right],\nonumber \\
4\pi S^{2}P_{r}a^{4} & = & -\frac{1}{2}\left[k\left(\left|R_{+}\right|^{2}+\left|R_{-}\right|^{2}\right)-\frac{(2l+1)}{\sin r}\frac{R_{+}^{\ast}R_{-}}{i}\right]\label{3.3}\\
 &  & \times\left(T_{+}^{\ast}T_{-}+T_{+}T_{-}^{\ast}\right),\nonumber \\
\mbox{and} &  & \text{}\nonumber \\
4\pi S^{2}\left(\rho-3P\right)a^{3} & = & -M\left(\left|Z_{+}\right|^{2}-\left|Z_{-}\right|^{2}\right).\nonumber 
\end{eqnarray}
 In the last equation for the trace, $3P=P_{r}+P_{\theta}+P_{\phi}$.
Differentiating the energy density with respect to $\ a$ we find
the energy conservation law as 
\begin{equation}
4\pi S^{2}\frac{d}{da}\rho a^{4}=-M\left(\left|Z_{+}\right|^{2}-\left|Z_{-}\right|^{2}\right)=4\pi S^{2}\left(\rho-3P\right)a^{3}.\label{3.4}
\end{equation}
WKB results for the quantities $\rho a^{4}\mbox{\,\,\ and\,\,}Ea^{4}=4\pi\intop_{0}^{r}\mbox{S}^{2}\rho a^{4}dr$
are shown in Fig. (\ref{energy}). 

Comparing Eq. (\ref{3.4}) with Eq. (\ref{3.1a}), it is important
to note that 

\begin{eqnarray*}
4\pi S^{2}na^{3} & = & \left|R_{-}T_{-}\right|^{2}+\left|R_{+}T_{+}\right|^{2}
\end{eqnarray*}

\begin{eqnarray*}
and\,\,\,4\pi S^{2}\frac{d(\rho a^{4})}{d(aM)} & = & \left|R_{-}T_{-}\right|^{2}-\left|R_{+}T_{+}\right|^{2}=4\pi S^{2}a^{3}\left(n_{-}-n_{+}\right)
\end{eqnarray*}

so that the rate of energy density change per unit mass is equal to
the difference in the comoving particle number density of the '-'
and '+' states.

From Eq. (\ref{3.4}), $d\left(\rho a^{3}\right)=-Pda^{3}$, which
is the Second Law of Thermodynamics for adiabatic expansion. This
equation can be processed to give the time independent entropy 
\begin{equation}
\mathcal{S}\left(r\right)=\frac{a^{3}\left(\rho+P\right)}{T}.\label{3.4a}
\end{equation}
 The total energy inside the closed Universe is

\begin{eqnarray}
Ea^{4} & = & \int\nolimits _{0}^{\pi}dr\int\rho a^{4}\sin^{2}rd\Omega\nonumber \\
 & = & -k\left(T_{+}^{\ast}T_{-}+T_{+}T_{-}^{\ast}\right)-Ma\left(\left|T_{+}\right|^{2}-\left|T_{-}\right|^{2}\right).\label{eq:Energy-Ea4}
\end{eqnarray}

Dividing this by the conserved comoving total number $Na^{3}=1$,
and using similar process with Eq. (\ref{3.4}), we get the equations
for the average comoving energy of one Dirac particle and its time
evolution: 
\begin{eqnarray}
\varepsilon a & = & -k\left(T_{+}^{\ast}T_{-}+T_{+}T_{-}^{\ast}\right)-Ma\left(\left|T_{+}\right|^{2}-\left|T_{-}\right|^{2}\right)\text{ and}\label{3.5}\\
\frac{d\varepsilon a}{da} & = & -Ma\left(\left|T_{+}\right|^{2}-\left|T_{-}\right|^{2}\right).\label{3.6}
\end{eqnarray}
 Eqs.(\ref{3.5} and \ref{3.6}) allow us to write 
\begin{eqnarray}
2M\left|T_{\pm}\right|^{2} & = & M\mp\frac{d\varepsilon a}{da}\text{ and}\label{3.7}\\
k\left(T_{+}^{\ast}T_{-}+T_{+}T_{-}^{\ast}\right) & = & a^{2}\frac{d\varepsilon}{da}.\label{3.8}
\end{eqnarray}
 We can also express the evolution of the moving energy in any Friedmann
type model by the equation 
\begin{equation}
\left(a^{\prime}\frac{d}{da}a^{\prime}\frac{d}{da}+4k^{2}+4M^{2}a^{2}\right)\frac{d\varepsilon a}{da}-4M^{2}a\left(\varepsilon a\right)=0.\label{3.9}
\end{equation}
Eq. (\ref{3.9}) can be solved, at least numerically or by series
method, for any behaviour of $a$.

\begin{figure}
\includegraphics{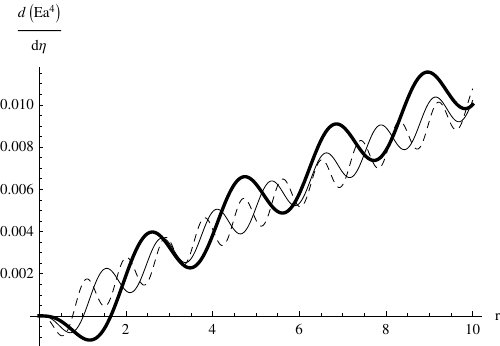}

\caption{\label{fig:Energy-flowRate}Comoving energy flow rate for flat universe
at a given time and mass. Thick solid, dashed and normal solid lines
are for k =3/2, 5/2 and 7/2 respectively. All are for $l$ = 1/2. }
\end{figure}

\begin{figure}
\includegraphics{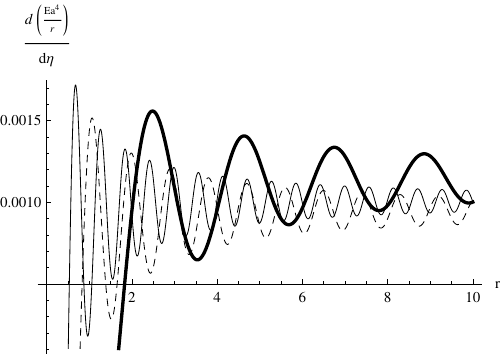}

\medskip{}

\includegraphics{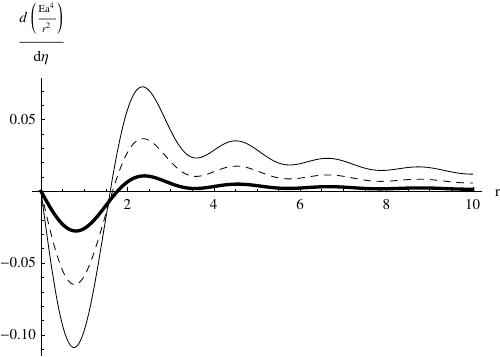}

\caption{\label{fig:Energy-flowRate-byr}Rate of flow of comoving energy for
the flat universe. The upper one is the energy divided by r and lower
one is the energy flowing out of a unit surface area at a distance
r. All are for $l$ = 1/2 with k =3/2, 5/2 and 7/2 represented by
thick solid, dashed and normal solid lines respectively }
\end{figure}

\begin{figure}
\includegraphics{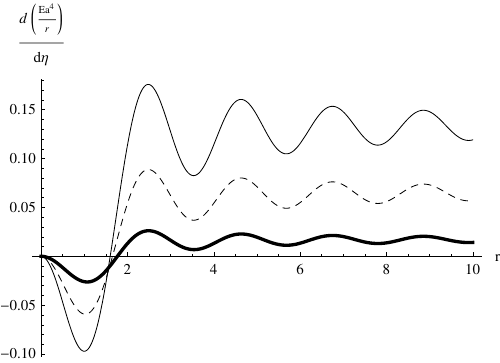}

\medskip{}

\includegraphics{Flat_EnergyCurrent_by_r2_variousM}

\caption{\label{fig:Energy-flowRate-M varied}As in the figure - \ref{fig:Energy-flowRate-byr}
for $l$ = 1/2 and k = 3/2 with different M. Thick solid, dashed and
normal solid lines are for M = 0.5, 1 and 1/5 respectively. }
\end{figure}

For the flat Universe, the energy density and rate of flow of energy
can be obtained explicitly as we did for particle number, by using
the expression for $R_{\pm}$~~given by Eq. (\ref{eq:R for flat})
and $T_{\pm}$~~given by Eq. (\ref{2.2}) in Eq. \ref{eq:Energy-Ea4}
. The plots of these for some representative cases have been shown
in Figs. (\ref{fig:Energy-flowRate}), (\ref{fig:Energy-flowRate-byr})
and (\ref{fig:Energy-flowRate-M varied}).

\begin{align}
\frac{1}{\dot{a}}\frac{\partial(Ea^{4})}{\partial\eta} & =\frac{\partial(Ea^{4})}{\partial a}=\left|A_{\pm}\right|^{2}\left[\frac{M}{2k}\left(\left|T_{+}\right|^{2}-1\right)r^{2}j_{l-1/2}j_{l+1/2}\right]\nonumber \\
+ & \left|A_{\pm}\right|^{2}\frac{M}{2}\left(1-2\left|T_{+}\right|^{2}\right)\intop_{0}^{\,\, r}\left(r\, j_{l+1/2}\right)^{2}dr
\end{align}

\section{Conclusions}

In this work, we have investigated the behaviour of the Dirac field
in closed FRW spacetime using NP formalism - spinor analysis. This
work has two important aspects. One is the method itself; we want
to illustrate the fact that the NP formalism is a very powerful and
convenient method to study the behaviour of spin fields in curved
background spacetime. It yields more tractable equations for the eigenfuctions
compared to the conventional tensorial method. It appears to be more
natural as it goes into very quantal nature of the fields as illustrated
by its ability to explain the absence of super-radiance of neutrinos
in Kerr spacetime\cite{Chandra}. Another aspect is the solution of
the field equations itself that have become more transparent in revealing
the nature of the distributions of the particles and energy. Ultimately
all matter is fermionic, so behaviour of Dirac field in expanding
Universe must reveal the unknown features of the Universe and also
unleash some facts about structure formation. 

While the free Maxwell and Dirac fields, in Friedmann-Robertson-Walker
spacetime were investigated using the Newman-Penrose method, all the
variables were found to be separable, and the angular solutions were
the spin-weighted spherical harmonics. The massless fields came to
have the usual exponential time dependencies. All the radial parts
reduced to the quantum mechanical barrier penetration problem, with
well behaved potentials that are basically the centrifugal energies.
The potentials seen by one component of the Dirac field, $R_{+}$,
are interesting; its lowest angular momentum state sees no potential
in the flat universe, while it sees an attractive one throughout the
open universe; from afar, all angular momentum states of this component
see attractive potential in the open universe. Consequences of this
effect may provide a means to determine whether the Universe is flat,
open or closed. All the radial equations are solved.

We have looked into the properties of the massive Dirac field in detail
for closed FRW spacetime. The radial dependence in the open Universe
is also the appropriate limit. The flat case can be considered to
be the limit of either the closed or open cases for $K\rightarrow0$.
Thus, in the complexified $r\mbox{-plane}$, the closed Universe lies
along the real axis, open along the imaginary and flat near the origin.
So, we may consider the proposition that, as an overall flat space
expands, it fragments into under-density voids that develop locally
as an open FRW space along the imaginary axis, and the overdensity
regions evolve locally into the large scale structures as the closed
counterpart. Hence the closed, open and flat Universes represent different
regions of a complexified space.

The radial dependence is the same for both the massless and massive
cases. The momentum is quantized in steps of unity from the minimum
value of $3/2$. The quantization condition for the comoving mementum
of a gravitationally trapped Dirac particle is \textit{k = l} + n
+ 1, where n$\geq0$ is an integer, and the total angular momentum
\textit{l} is a half integer.

The temporal solution (Eq. \ref{2.5}) gives us an important information,
viz,. at late times of the evolution / expansion of the Universe when
$a\rightarrow\infty,\,\varepsilon_{0}\rightarrow M$, so $T_{+}\rightarrow0$
and $T_{-}\rightarrow-1$, showing that one of the solutions decays
in time as the Universe expands, while the other solution grows. This
leads us to conclude that the former one may be the negative energy
solution while the latter is the positive energy solution. This will
be confirmed in future work on the representation of the Dirac matrices.
 The variations in number densities exist from the very initial time
of the big-bang. There was indication from our early work \cite{uk1}
that, although the comoving number density of the massless field is
locally conserved at every spacetime point, it is not so for the massive
case. Even though the Universe starts out with equal number of particles
with $\left|T_{+}\right|^{2}=\left|T_{-}\right|^{2}=1/2$, at late
times the former goes down towards zero while the latter grows to
unity keeping the sum conserved and equal to unity. As the universe
expands and the particles lose their kinetic energy, the time dependence
shows that the $\left|Z_{+}T_{+}\right|^{2}$ representing the number
of particles in this model will decay while $\left|Z_{-}T_{-}\right|^{2}$
grows. This may be mimicking the behavior of the massive Dirac field
in Minkowski spacetime where the negative energy solutions exist only
in the relativistic regime while the positive energy solution dominates
in the non-relativistic regime.\cite{Roman-1965}. This could be the
source of the local particle current that exists at every point in
the FRW spacetime. Nonetheless, the total comoving number within the
whole volume is still conserved. Further, the currents are set up
in such a way that the over-densities are further enhanced while the
under-densities get more depleted as the Universe expands. 

\begin{figure}
\includegraphics[scale=1.5]{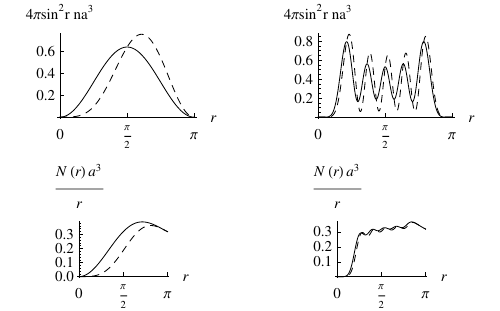} \\
\caption{The number of Dirac particles in closed FRW space-time. The upper
two are the local surface densities of particle number, and the lower
two are the total number divided by $r$. On the left are $k=3/2,\, l=1/2$,
and on the right are $k=17/2,\, l=7/2$. Solid curves are the respective
quantities at the instance of the big bang when $a=0$, and the dashed
at $a=10$. It is seen that as the Universe expands, the density contrast
become more enhanced. The lower graphs show that the particle distribution
has the same behaviour as that demanded by the flat rotation curves
of galaxies. }

\label{numbercl} 
\end{figure}

\begin{figure}
\includegraphics[scale=1.5]{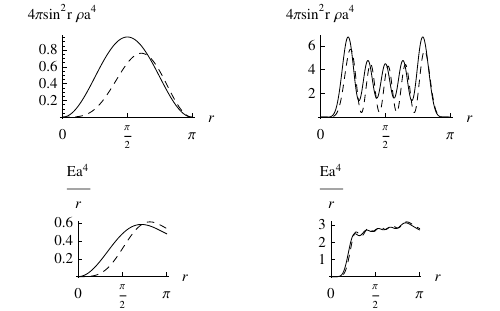} \\
\caption{The energy of the Dirac field for the same values of $k\mbox{ and }l$
as in Fig. (\ref{numbercl}). But here the values are scaled for $a=0\mbox{ and }a=10$
to appear to be almost equal so that they can be compared. The energy
follows the particle number as it should.}

\label{energy} 
\end{figure}

The energy distribution also follows the number distribution. We find
that the number and energy distributions that are setup have exactly
the same form as required by the flat ratation curves of galaxies.
The Newtonian gravitational potential that is required for the flat
rotation curve is that the square of the rotational velocity of the
particles, $v^{2}\propto\frac{M\left(r\right)}{r}\propto\frac{N\left(r\right)}{r}=\mbox{constant}$.
Our results also show that the plots of the quantities $\frac{E\left(r\right)}{r}\mbox{ and }\frac{N\left(r\right)}{r}$
become flat at a short distance from the origin as the Figs. (\ref{numbercl}
and \ref{energy}) show. Even the surface densities of particles and
energy at a distance r show the same nature. The particle number and
energy distributions have also been plotted for different masses and
different conformal times (in the units of Hubble time) and it is
seen that density contrast is enhanced for larger mass M as well as
with the elapse of time as expected.

The over- and under-density fluctuations that exist from the very
begining further get enhanced with the expansion of the Universe.
As a consequence, fragmentation can be induced as the underdensity
regions inbetween the over-densities become highly depleted. The angular-momentum
contained by these structures must also be the source of rotation
of galaxies and clusters. As our Universe is almost flat, any gravitationaly
bound large scale structure will mimic a closed Universe. 

This work has to be continued further, and these preliminary results
confirmed by simulation. Also, quantization of the Dirac field must
be done and the corrections accounted for conclusive arguments. Another
extension would be the study of quantum electrodynamics of the interacting
Dirac and Maxwell fields.

\pagebreak{}


\begin{figure}
\includegraphics{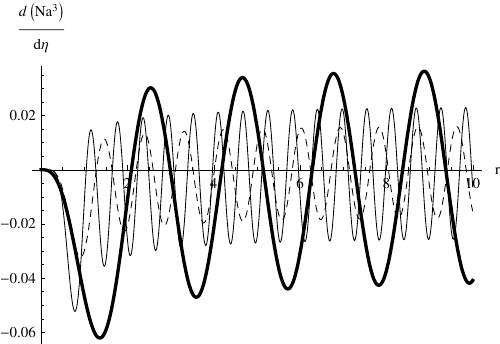}

\medskip{}

\includegraphics{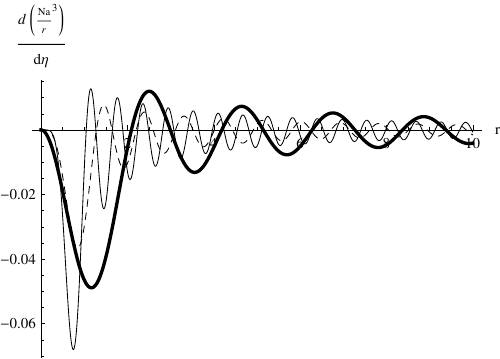}

\caption{\label{fig:TimeRate-Na3}The comoving particle current for given mass
and conformal time in flat FRW space time. The thick solid curve is
for \textit{l} = 1/2 and k = 3/2; the dashed curve is for \textit{l}
= 3/2 and k = 7/2; and normal solid curve is for \textit{l} = 5/2
and k = 11/2. The lower one is the particle number divided by r. The
later graphs show that the particle distribution has the same behaviour
as that demanded by the flat rotation curves of galaxies. }
\end{figure}

\begin{figure}
\includegraphics{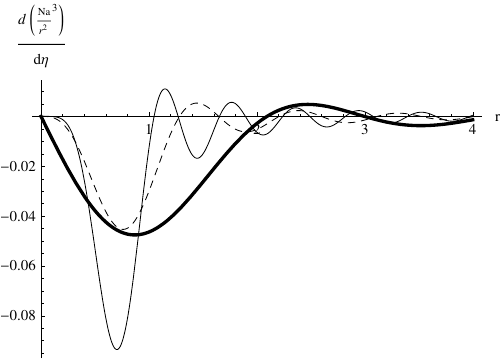}

\caption{\label{fig:TimeRate-Na3byr2}The rate of flow of comoving particles
through a unit surface at r in flat FRW space time. The thick solid
curve is for \textit{l} = 1/2 and k = 3/2; the dashed curve is for
\textit{l} = 3/2 and k = 7/2; and normal solid curve is for \textit{l}
= 5/2 and k = 11/2.}
\end{figure}

\end{document}